# Kinetic Modeling of Transient Electroluminescence reveals TTA as Efficiency-Limiting Process in Exciplex-Based TADF OLEDs

Jeannine Grüne[1], Nikolai Bunzmann[1], Moritz Meinecke[1],
Vladimir Dyakonov[1], Andreas Sperlich[1*]

[1]Experimental Physics 6, Julius Maximilian University of Würzburg, Am Hubland,
97074 Würzburg, Germany

ABSTRACT

Organic light emitting diodes (OLEDs) based on thermally activated delayed fluorescence (TADF) show increased efficiencies due to efficient upconversion of non-emissive triplet states to emissive singlets states via reverse intersystem crossing (RISC). To assess the influence of the characteristic efficiency-enhancing RISC process as well as possible efficiency-limiting effects in operational OLEDs, we performed temperature-dependent measurements of transient electroluminescence (trEL). With kinetic modeling, we quantify and separate the impact of different temperature-dependent depopulation processes and contributions to EL in the established donor:acceptor model system m-MTDATA:3TPYMB. The underlying rate equations adapted for EL measurements on TADF systems include radiative and non-radiative first- and second-order effects. In this way, we are able to evaluate the non-radiative recombination and annihilation processes with respect to their efficiency-limiting effects on these OLEDs. On the one hand, we evaluate the depopulation of intermolecular exciplex triplet states via non-radiative direct triplet decay, RISC and triplet-triplet annihilation (TTA). On the other hand, we determine the contribution to EL from the formation of singlet exciplex states via polarons, RISC and TTA. Our results show that TTA accounts for a significant part to triplet depopulation and contributes to EL while limiting the overall device quantum efficiency.

INTRODUCTION

Organic light emitting diodes (OLEDs) based on thermally activated delayed fluorescence (TADF) have attracted much attention since their first report[1], due to their ability to achieve 100% internal quantum efficiency. The underlying mechanism is based on harvesting the non-emissive triplet states and up-convert them to emissive singlet states via reverse intersystem crossing (RISC).[2,3] Since this mechanism is driven by thermal energy, a necessary molecular property for this purpose is a small energy gap $\Delta E_{ST}$ between the first excited singlet and triplet states, e.g. by minimizing the overlap of involved HOMO and LUMO orbitals.[4] Several promising concepts are under consideration to realize efficient TADF. Either by combining donor and acceptor molecules that can share an intermolecular exciton at their interface, also called exciplex[5,6] or alternatively by designing molecules that incorporate donor and acceptor moieties, enabling intramolecular excitons[7,8]. However, a small energy gap $\Delta E_{ST}$ is not the only requirement for efficient TADF devices.[9] The efficiency-increasing RISC rate must be large enough to outperform efficiency-reducing non-radiative recombination and annihilation processes.

This work focuses on a model system for the investigation of exciplex-based OLEDs employing the donor 4,4',4"-Tris[(3-methylphenyl)phenylamino]triphenylamine (m-MTDATA) and the acceptor Tris(2,4,6-trimethyl-3-(pyridin-3-yl)phenyl)borane (3TPYMB). This material combination still attracts a lot of attention in literature[10-12], however, according to reports, it never achieved an external quantum efficiency (EQE) higher than 12.9%[12]. Since intermolecular donor:acceptor systems are generally characterized by a small energy gap $\Delta E_{ST}$, the limitation in efficiency is surprising. On the one hand, a comparatively low photoluminescence quantum yield (PLQY) of 45%[11] can indicate a lower EQE. However, exciplex states are non-absorbing[13], thus requiring a previous optical excitation of the

---

donor/acceptor molecules. Intermediate transitions to exciplex states and associated loss mechanisms on the pristine molecules can lower the PLQY without involvement of the exciplex states. On the other hand, electrical excitation generates exciplex states directly, but second-order loss mechanisms, such as annihilation effects, can reduce EQE significantly. To investigate the influence of the characteristic RISC process as well as possible efficiency-limiting effects on device quantum efficiency, we carried out temperature-dependent transient electroluminescence (trEL) measurements. In order to identify loss processes, we analyzed these transient measurements with a suitable kinetic model adapted for EL measurements on TADF devices. The underlying rate equations for singlets, triplets and polarons include second-order terms to consider annihilation processes.

We aim to address all relevant processes involved in an operating OLED device, since previous reports refer either to modeling of transient photoluminescence (trPL) on TADF materials[9,14,15], transient EL of TADF devices without considering annihilation effects[16] or on transient EL of non-TADF devices considering for second-order processes[17-20]. However, as shown in the following, it is especially important in TADF devices to account for annihilation effects. On the one hand, annihilation effects are often based on interactions with triplet excitons and are therefore more influential with increasing triplet density in the device. However, the RISC process positively counteracts depopulation via annihilation as it reduces the triplet density. Thus, the magnitude of the RISC rate and the effective triplet density are important factors determining the influence of efficiency-limiting processes.[21] On the other hand, particularly in electrically driven devices, the consideration of second-order effects is important since significantly more triplet excitons are generated through injection than in optical excitation. Due to their long lifetime, triplet excitons accumulate which facilitates the influence of annihilation processes.[22] However, a difficulty in transient EL modeling is that the OLED first reaches an equilibrium state, determined by all present rates, before the transient EL decay is recorded at the time of switching off the applied voltage. This results in a population ratio of triplets to singlets in the operational steady-state of up to 67:1, differing significantly from the generation ratio of 3:1. Therefore, a suitable kinetic model has to determine the population number in the operational steady-state as starting parameters for the transient EL decays.

In this paper, we show how adequate kinetic modeling of temperature-dependent transient EL measurements helps to quantify efficiency-increasing and -limiting first- and second-order processes. We demonstrate a rate equation based fit procedure, which accurately reproduces transient EL measurements and provides required first- and second-order rates. The procedure is especially advantageous for exciplex-based systems, since rate extraction by transient PL is significantly complicated due to non-absorbing exciplex states. Ultimately, the results enable us to quantify the impact of efficiency-limiting processes in operational devices and to determine the contribution of relevant processes to the steady-state EL.

THEORETICAL

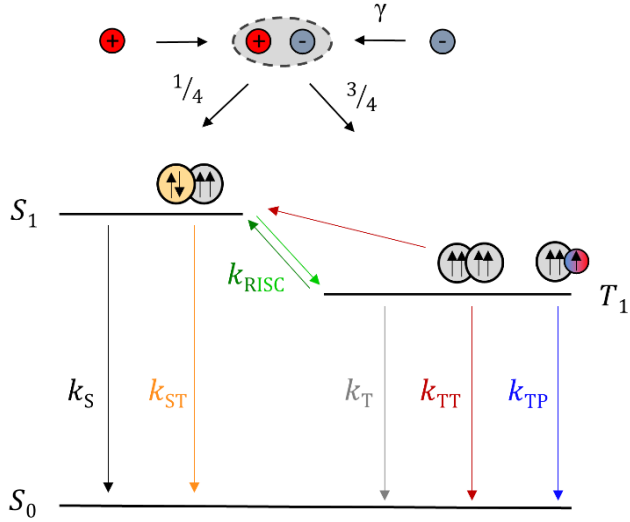

**Figure 1.** Kinetic model of a three-level system for transient EL. Injected charge carriers form excitons with the Langevin recombination rate γ. Excitons populate singlet and triplet states according to 1:3 spin statistics. Possible first-order processes include singlet decay rate $k_S$, (reverse) intersystem crossing rate $k_{(R)ISC}$ and non-radiative decay rate $k_T$. Second-order processes are considered with singlet-triplet annihilation rate $k_{ST}$, triplet-triplet annihilation rate $k_{TT}$ and triplet-polaron annihilation rate $k_{TP}$.

**Kinetic Model. Figure 1** schematically illustrates the kinetic model including all of the considered processes for the analysis of electrically driven TADF OLEDs. In EL measurements, the excitation is given by electrical injection. The injected charge carriers form excitons assuming Langevin recombination with rate γ, populating continuously the singlet states $S_1$ and triplet states $T_1$ in a 1:3 ratio according to spin statistics.[23,24] Afterwards, the excitons undergo transitions with rate-dependent probabilities. After a certain time of continuous OLED operation, an equilibrium of singlet, triplet and charge carrier densities, determined by all present rates, is achieved. At this point, the OLED has reached the steady-state of EL operation. The first-order processes in this kinetic model include the decay rate $k_S$ of the $S_1$ states, the direct intersystem crossing (ISC) rate $k_{ISC}$ as well as the reverse ISC rate $k_{RISC}$ for transitions between $S_1$ and $T_1$ and the non-radiative decay rate $k_T$ of the $T_1$ states. For the sake of simplicity, the singlet decay rate $k_S$ considers both, the radiative and non-radiative decay.[19,20] Density-dependent second-order processes involve all relevant annihilation effects that are considered in literature in relation to TADF.[10,12,25,26] These are particularly effects that involve interactions with triplet excitons. Regarding the long lifetime of triplet excitons, the probability of annihilation processes is expected to be higher for triplet excitons than for singlet excitons.[22] Furthermore, the triplet density in electrically driven devices is significantly higher than in optical excitation due to their generation ratio. Therefore, we consider the rates for triplet-triplet annihilation $k_{TT}$, triplet-polaron annihilation $k_{TP}$ and singlet-triplet annihilation $k_{ST}$.

Triplet-triplet annihilation (TTA) is based on the interaction of two triplet excitons. Their encounter leads to an intermediate compound or scatter state $(T_1T_1)$ which, according to spin statistics, can be transformed into a singlet, triplet or quintet state.[27,28] Since no triplet exciton is lost for the quintet state, the rate $k_{TT}$ can effectively be transformed into 1/4 for the singlet and 3/4 for the triplet states (see Supporting Information):

$$T_1 + T_1 \longrightarrow \begin{cases} \xrightarrow{5/9} T_1 + T_1 \\ \xrightarrow{3/9} T_1 + S_0 \\ \xrightarrow{1/9} S_1 + S_0 \end{cases} \xrightarrow{\text{Supp. Inf.}} \begin{cases} \xrightarrow{\frac{3}{4}k_{\text{TT}}} T_1 + S_0 \\ \xrightarrow{\frac{1}{4}k_{\text{TT}}} S_1 + S_0 \end{cases} \tag{1}$$

Triplet-polaron annihilation (TPA) is mainly a Förster-type transfer and can be described as an annihilation process of triplet excitons with the spin state of the polaron, resulting in the transfer of the polaron in the ground state $n$ to an excited state $n^*$:[19,22,29]

$$T_1 + n \xrightarrow{k_{\text{TP}}} S_0 + n^* \tag{2}$$

Based on the long-living triplet states, singlet-triplet annihilation (STA) can also occur at high triplet densities.[20] STA is a spin-allowed Förster-type energy transfer, whereby the triplet exciton will be raised to a higher triplet state $T_n$ with relaxation of the singlet exciton into the ground state $S_0$:[19]

$$S_1 + T_1 \xrightarrow{k_{\text{ST}}} S_0 + T_n \tag{3}$$

Considering these second-order effects in combination with the linear first-order processes mentioned above, the kinetics of the excited states in an operating OLED can be described with the following rate equations. We assume a polaron density $n$ due to injected current density $j$, which recombines with the Langevin recombination rate $\gamma$ to excitons:[19,20,22,30]

$$\frac{dn}{dt} = \frac{j}{ed} - \gamma n^2 \tag{4}$$

While $j/ed$ describes the polaron generation within the emissive layer thickness $d$ and elementary charge $e$, the second term $\gamma n^2$ subtracts the formed excitons.[17,29] The Langevin recombination rate is given by $\gamma = e(\mu_e + \mu_h)/(\varepsilon_0 \varepsilon_r)$[18,20] with $\mu_e$ and $\mu_h$ representing the mobilities of electrons and holes in the emission layer[10,31] and $\varepsilon_0$ and $\varepsilon_r$ the vacuum and emission layer permittivity, respectively (see Supporting Information).

A quarter of the created excitons ($\frac{1}{4}\gamma n^2$) directly contributes to population of the singlet states. The first-order processes populating and depopulating the singlet states are intersystem crossing ($k_{\text{ISC}}S_1$), reverse intersystem crossing ($k_{\text{RISC}}T_1$) and the singlet decay ($k_S S_1$). Based on equations 1 and 3, the annihilation processes affecting the singlet states are triplet-triplet annihilation ($\frac{1}{4}k_{\text{TT}}T_1^2$) and singlet-triplet annihilation ($k_{\text{ST}}T_1S_1$). The resulting rate equation for the kinetics of the singlet states therefore follows:

$$\frac{dS_1}{dt} = -k_S S_1 - k_{\text{ISC}}S_1 + k_{\text{RISC}}T_1 + \frac{1}{4}k_{\text{TT}}T_1^2 - k_{\text{ST}}T_1 S_1 + \frac{1}{4}\gamma n^2 \tag{5}$$

The triplet states are directly populated by the other three quarters of initial excitons ($\frac{3}{4}\gamma n^2$). The first-order processes are analogous to the singlet state. Annihilation processes interfering with the triplet states are triplet-triplet annihilation ($\frac{5}{4}k_{\text{TT}}T_1^2$) and triplet-polaron annihilation ($2k_{\text{TP}}T_1 n$), based on equation 1 and 2. The resulting rate equation for the kinetics of the triplet states is therefore:

$$\frac{dT_1}{dt} = -k_T T_1 + k_{\text{ISC}}S_1 - k_{\text{RISC}}T_1 - \frac{5}{4}k_{\text{TT}}T_1^2 - 2k_{\text{TP}}T_1 n + \frac{3}{4}\gamma n^2 \tag{6}$$

The pre-factor for TTA $\left(\frac{1}{4} \text{ or } \frac{5}{4}\right)$ is obtained from the different cases of equation 1 (see Supporting Information). The pre-factor for TPA (2) is based on the assumption of balanced electron and hole density in the emissive layer, which is discussed in more detail below.

METHODS

**OLED Fabrication.** The OLEDs were fabricated on indium-tin-oxide (ITO) pre-coated glass substrates (1.62 cm$^2$). A 40 nm hole injection layer of poly(3,4-ethylendioxythiophene):polystyrolsulfonate (PEDOT:PSS, 4083Ai) was spin-coated and subsequently annealed for 10 min at 130 °C. In a vacuum thermal evaporation chamber with base pressure $p < 10^{-6}$ mbar, the donor m-MTDATA and the acceptor 3TPYMB as well as the calcium (Ca) and aluminum (Al) cathode were deposited by thermal evaporation. The resulting device structure is ITO/PEDOT:PSS/m-MTDATA(30nm)/ m-MTDATA:3TPYMB(70nm,1:1)/3TPYMB(30nm)/Ca(5nm)/Al(120nm).

**Transient EL Measurements.** Transient EL measurements were performed in a nitrogen flow cryostat (Oxford 935). A function generator (Agilent Technologies 81150A) was used for pulsed OLED operation and the luminescence was detected with a photodiode (Hamamatsu Si photodiode S2387-66R). A current-voltage amplifier (Femto DHPCA-100) amplified the light-induced photocurrent, which was recorded with a digitizer card (GaGe Razor Express 1622 CompuScope). OLEDs were driven at a current density of $10 \frac{\text{mA}}{\text{cm}^2}$ for 4 ms to ensure steady-state operation. The response time of the setup is determined by the bandwidth of the current-voltage amplifier of 1 MHz at a gain level of $10^6$ V/A. To receive a defined sample data count, the step-size of the transient EL measurements was interpolated to this time resolution.

**Fit Method.** To fit the measured transient EL data, we used the `optimize.curve_fit()` function from the *SciPy* library (v. 1.1.0.) in Python (v. 3.7.6), using the numerical solution of rate equations 4 – 6 as a fitting function. The procedure differs from fitting transient PL traces after pulsed laser excitation, since the steady-state of the OLED must first be simulated based on the rate equations 4 – 6 and the injected current density. The knowledge of the population densities in the steady-state is important because it determines the initial state of the transient decay. Since this population ratio is determined by an equilibrium of all present rates, it differs from the generation ratio of 1:3. For the transient decay, the current density $j$ in equation 4 is set to zero, which is equivalent to switching off the OLED. In order to determine the required rates from the transient EL decays, an iterative numerical solution of the rate equations is fitted to the measured EL transients starting from 10 μs after switching off the OLED (to avoid influence of electrical artifacts). The source code as well as more details about the fitting procedure are given in the Supporting Information.

RESULTS AND DISCUSSION

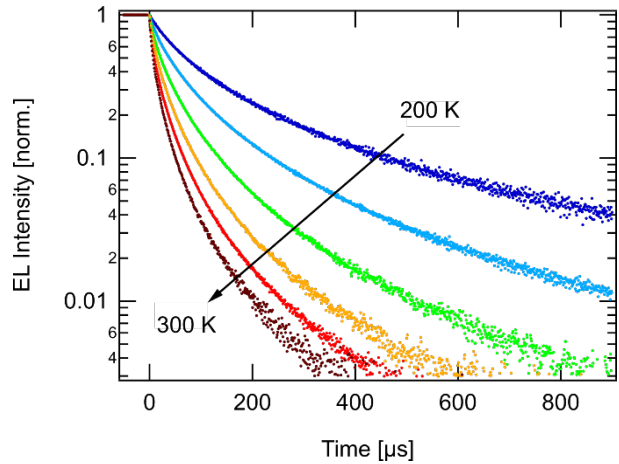

**Figure 2.** Temperature-dependent transient EL decays for an OLED with emission layer of m-MTDATA:3TPYMB. The traces show a temperature-activated non-linear behavior.

**Experimental Results. Figure 2** shows transient EL measurements of the reported system in semi-logarithmic representation for temperatures from 200 K to 300 K in 20 K steps (corresponding temperature-dependent EL spectra are shown in Figure S1). On the one hand, we observe a temperature-activated process since the transient EL decays become faster with increasing temperature. On the other hand, the shape of the transients deviates from the expected biexponential decay by prompt and delayed fluorescence. A deviation from linear traces in semi-logarithmic presentation indicates the influence of second-order processes, e.g. annihilation processes. Murawski et al. and Baldo et al. reported deviations from mono-exponential decay in transient EL measurements of phosphorescent OLEDs and both attributed this effect to TTA.[17,22] However, in order to consider also the other annihilation effects mentioned, we used the rate equations 4 – 6 including STA, TPA and TTA to find the dominant second-order process responsible for the shape of the traces in Figure 2.

Before fitting the transient EL decays from Figure 2, we first identify the dominant annihilation effect that determines the shape of the transients. Therefore, we use reduced rate equations of equation 4 – 6 with one annihilation effect, respectively and investigate its influence on the numerical solutions.

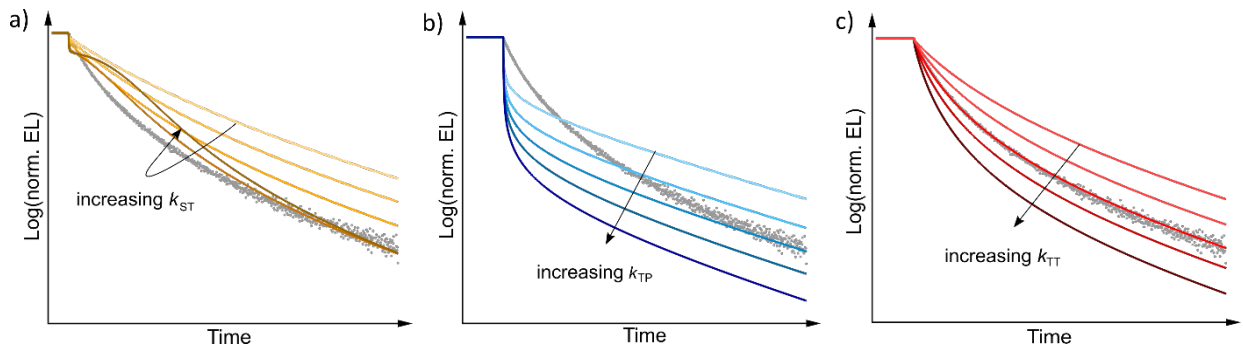

**Figure 3.** Influence of different annihilation processes on the transient decay. Colored transients show the numerical solutions of reduced rate equations 4 – 6 with varying annihilation rates of a) singlet-triplet annihilation, b) triplet-polaron annihilation, c) triplet-triplet annihilation. The gray curve always displays the experimental EL transient recorded at 200 K. Only a dominant TTA mechanism can reproduce the experimental transient EL curve accurately.

To visualize the influence of the different annihilation effects on the transient EL decay, **Figure 3** shows the numerical solution of the reduced rate equations 4 - 6 with the variation of one annihilation rate, respectively. Figure 3a presents simulated transient EL traces with different STA rates. In Figure 3b, the TPA rate was varied and Figure 3c shows the influence of different TTA rates. For comparison with experimental data, the gray curve always displays the experimental EL transient recorded at 200 K. The fitting method described in the method section provided only reasonable fits for the case of TTA as annihilation effect. Because STA and TPA alone cannot reproduce the experimental data, suitable values for the annihilation rates had to be set manually for illustrative purpose.

Since the influence of STA depends on the product of singlet and triplet population densities, STA does not have a relevant influence due to the smaller density and shorter lifetime of singlet excitons. At lower values of STA, the traces are still approximately linear in the semi-logarithmic presentation. At higher orders of magnitude, the influence can be observed mainly at the beginning of the transient. Overall, this annihilation effect cannot reproduce the shape of the measured trace. Furthermore, the measured transient EL decays are current dependent (Figure S2), which contradicts the dominance of STA.[22] The influence of the TPA rate on EL transients (Figure 3b) is particularly strong shortly after switching off the OLED, i.e. in the first microseconds. The reason for this is the finite transit time of polarons in the device active layer after the voltage turn-off, determining the time scale in which TPA can occur. After this time, the traces follow a mono-exponential decay. In OLEDs, the presence of majority charge carriers is often discussed, as they can promote TPA while minority charge carriers are used for exciton formation.[29] However, the influence of these majority carriers would also be negligible after the transit time of a few microseconds and can therefore not be responsible for the shape of the transient at later times. In contrast, TTA is present during a longer time span of the decay. Certainly, the influence of TTA is higher at the beginning of the transient, since most triplet excitons are present there. Nevertheless, TTA influences the EL decay as long as triplet excitons are existent, therefore the influence of TTA is present over the whole decay. Figure 3c confirms that considering TTA in the rate equations 4 – 6 matches the measured transient EL trace almost perfectly.

By comparing the behaviors of the modeled traces, we can observe that TTA is the mechanism, which is responsible for the characteristic shape of the transient EL traces. On the one hand, this result is consistent with the already mentioned statements of Murawski et al. and Baldo et al., who explained deviations in EL transients from linear progressions with TTA.[17,22] On the other hand, we carried out excitation power-dependent PL measurements (Figure S3). These measurements give the order of the process from the slope when plotted in a double logarithmic representation. Even though under optical excitation (where the triplet density is slightly lower), TTA can be verified by a slope of almost two[32], which implies the necessity of two triplet excitons for a radiative decay.

**Fit Results.** The results in the following section were obtained with the fit procedure introduced in the methods section, based on reduced rate equations 4 – 6 considering the annihilation effect TTA. In order to avoid distorting the fit by too many free parameters, the singlet decay rate $k_S$ is set to values according to literature[33]. Therefore, the fit procedure determined the ISC rate $k_{ISC}$, the RISC rate $k_{RISC}$, the non-radiative decay rate $k_T$ and the TTA rate $k_{TT}$, whereby $k_{ISC}$ is assumed to be temperature-independent[2]. Typically, transient PL measurements[9,14] or sometimes also PL quantum yield measurements[34] provide fundamental rates such as $k_S$, $k_{(R)ISC}$ and $k_T$. However, as already mentioned in the introduction, we are investigating an exciplex system whose CT/exciplex states are non-absorbent (Figure S4) and therefore not directly optically excitable.[35] Exciplex systems are defined by linear combination of the excited states of the donor and acceptor molecules whereby optical excitation can only generates an intramolecular exciton on the donor or acceptor molecule.[13] The subsequent charge transfer creates an intermolecular exciton at the interface (CT/exciplex state). Considering the initial excitation via an intramolecular excitation complicates the rate equations for trPL in contrast to directly absorbing CT states. However, electrical excitation (EL measurements) populates exciplex states directly, which simplifies the system to a three-level system. Therefore, transient EL measurements are particularly appropriate (compared to PL) for the determination of the RISC rate in these exciplex systems.

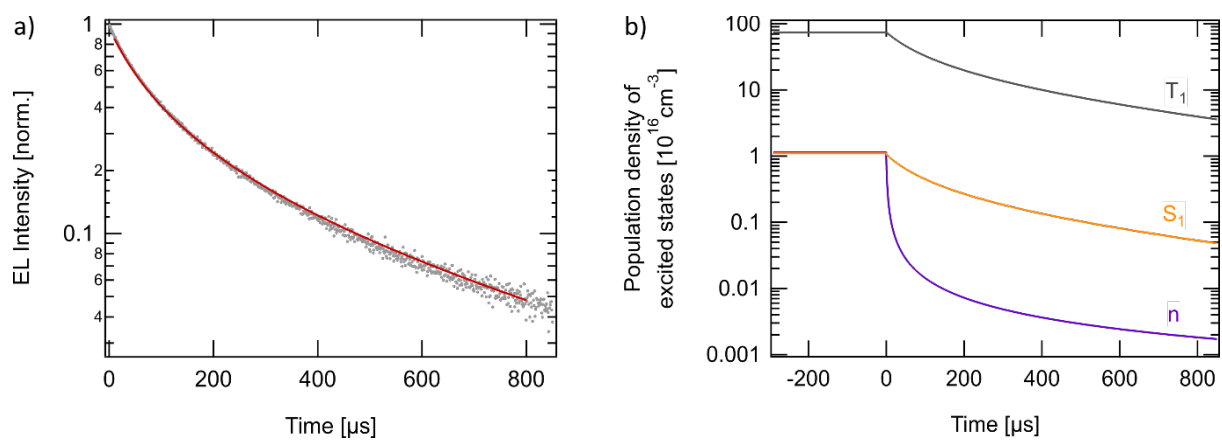

**Figure 4.** Results of the fit procedure for 200 K. a) Fit (red) of the iterative numerical solution of equations 4 – 6 with considering TTA as the dominant annihilation mechanism. b) Time-dependent population density of polarons n (purple), singlet states $S_1$ (orange) and triplet states $T_1$ (gray).

**Figure 4a** shows the resulting fit (red) of the transient EL measurement at 200 K. The RISC rate was determined to be $k_{RISC} = 4.7 \times 10^4$ s$^{-1}$ which corresponds to a RISC time of about 21 µs. The order of magnitude of this rate is consistent with reports from this or similar donor:acceptor TADF systems.[2,33] At the same time, the RISC rates in these intermolecular exciplex systems are smaller than for state-of-the-art intramolecular TADF emitters. Hence, the depopulation of triplet states by RISC proceeds slower and the long-living triplet excitons accumulate, enhancing the influence of TTA.[22] From the fit we determined a TTA rate of $k_{TT} = 1.4 \times 10^{-14}$ cm$^3$ s$^{-1}$. Baldo et. al., Kasemann et. al. and Murawski et. al. reported TTA rates from transient EL measurements on phosphorescent OLEDs of the same orders of magnitude.[17,20,22] Niwa et. al. determined a TTA rate of about $10^{-18}$ cm$^3$ s$^{-1}$ at helium temperatures by steady-state PL measurements of intramolecular TADF emitters.[26] Since the TTA rate is highly temperature dependent (Figure S5), the fitted TTA rates agree very well with previously discussed rates. The ISC rate was determined to be $k_{ISC} = 3.4 \times 10^6$ s$^{-1}$ which is consistent with reports for this material system.[2,33] The non-radiative decay rate $k_T$ is with a value of $k_T = 1.0 \times 10^2$ s$^{-1}$ about two orders of magnitude smaller than the RISC rate, implying that the long EL decay is not effectively shortened by non-radiative triplet decay. Therefore, the depopulation by non-radiative triplet decay plays a minor role, which will be discussed in more detail below. The fit curves for all temperatures (Figure S6) as well as the rates determined by the fit (Table S1) are available in the Supporting Information.

To quantify the density and lifetime of the excited states, **Figure 4b** shows the time-dependent population density of polarons (n), first excited singlet states ($S_1$) and first excited triplet states ($T_1$) slightly before and after switching off the OLED at time t = 0 µs for 200 K. The simulations for the population density are based on the fit results in Figure 4a and the rate equations 4 – 6. On the one hand, we observe that the polarons disappear within a few microseconds. This observation agrees well with the results shown in Figure 3b, suggesting that TPA only affects the beginning of the EL transients as there are negligibly few free polarons available in the device after several microseconds. On the other hand, Figure 4b shows that the steady-state triplet density (t < 0 µs) is significantly higher than the singlet density (67:1), facilitating annihilation processes such as TTA[21]. The actual triplet-to-singlet ratio when the OLED is switched off therefore does not match the electrical occupation of the states with the ratio of 3:1 according to simple spin statistics. In fluorescent OLEDs, when only considering the non-radiative decay rate, Shinar et al. showed that the triplet density can be up to $10^5$ times higher than the singlet density.[36] However, they assumed a relatively high ratio of triplet-to-singlet lifetime of $10^5$:1 and included no depopulation of the triplet states by annihilation processes or RISC mechanism in their estimation. Since the ratio of triplet-to-singlet lifetime in our system is smaller and RISC and TTA processes both play a significant role in the depopulation of triplet states, the triplet-to-singlet population ratio is reduced.

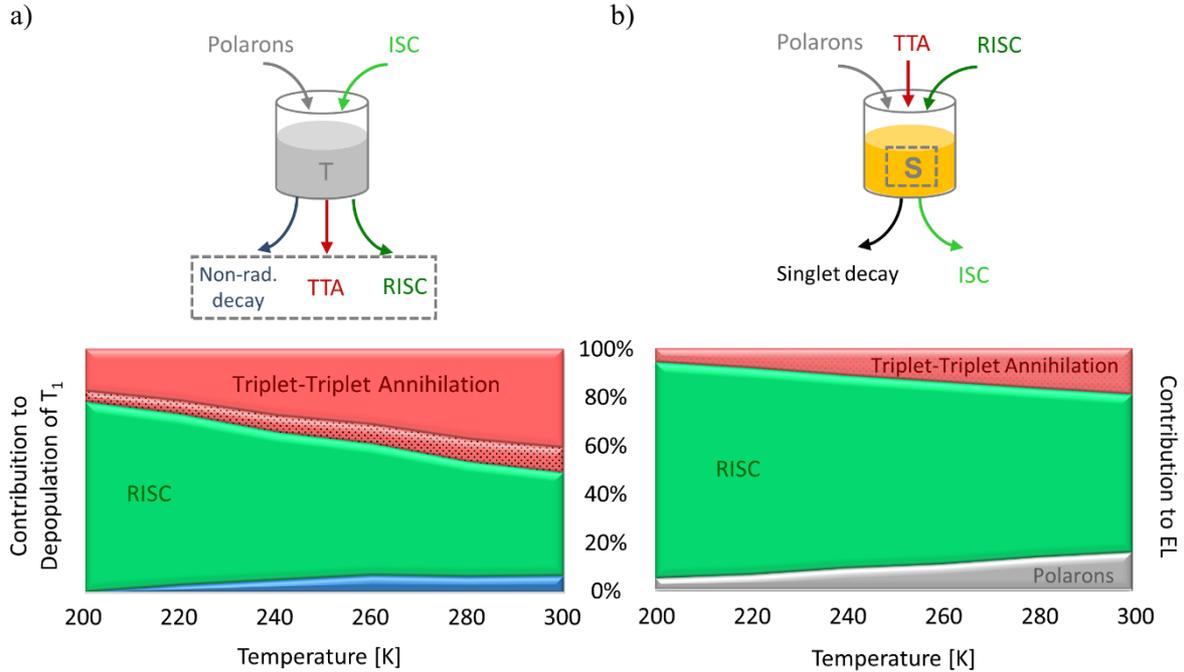

**Figure 5.** Schematic bucket model (top) and temperature-dependent impact (bottom) of different (de-)population processes. **a**) Contribution to depopulation of triplet state $T_1$ via non-radiative decay (blue), RISC (green) and TTA with creation of a singlet state (red, dotted) or a triplet state (red). **b**) Contributions to EL of three different processes: singlet excitons formed via polarons (gray), RISC (green) and TTA (red, dotted).

**Figure 5a** shows a scheme of the processes that contribute to the population (polarons, ISC) and depopulation (non-radiative decay, TTA, RISC) of the triplet states. To assess the influence of the processes depopulating the triplet states, we investigate the impact of the respective depopulation channels (gray framed rates in Figure 5a). Therefore, the diagram in Figure 5a represents the temperature-dependent contributions of the individual depopulation processes of the triplet states in operational conditions (added up to 100%). The data sets are derived from the fit results of the measured EL transients (Table S1), which we used to calculate the steady-state with equations 4 – 6. The first-order non-radiative decay process of the triplet states (blue) slightly increases with temperature but overall plays a minor role in the depopulation of the triplet states. Therefore, TTA and RISC are primarily responsible for the depopulation of the triplet states. Our fit procedure delivers a singlet-triplet energy gap of $\Delta E_{ST} = 23$ meV, which is comparable with previous reports for this material system.[11,33] The RISC rate is assumed to increase exponentially with temperature according to a Boltzmann factor. However, the impact of RISC on depopulation of triplet states $T_1$ decreases with increasing temperature. The reason for the decline is the proportion of TTA, which is also temperature-activated. An excess of injected charge carries cannot cause this effect, since the measurements are performed at the same current density for each temperature. Actually, in contrast to the RISC rate which is increasing following a Boltzmann factor $\exp(-c_1/T)$, the TTA rate increases almost exponentially with $\exp(c_2 \cdot T)$ (Figure S5). Consequently, the proportion of TTA in triplet density depopulation increases with temperature until it outperforms depopulation via RISC at room temperature. This observation is remarkable since at relatively high temperatures, the triplet lifetime is significantly reduced by the fast phonon-assisted decay, which normally overcomes TTA.[37,38] The fact that the proportion of TTA nevertheless increases indicates the enormous influence of TTA in this system. TTA has already been detected in other TADF systems.[26,39] However, the proportion of TTA in exciplex-based OLEDs is significantly higher. On the one hand, one factor is the investigation of operational OLEDs, since a higher triplet density is present in EL in contrast to PL. On the other hand, and this is the larger influence, the moderate RISC rate in exciplex-based systems allows an accumulation of triplet excitons. As a result, about 50% of the triplet depopulation is caused by TTA and only 44% by the - for TADF specific - RISC mechanism at room temperature. Since one TTA event results in the loss of at least one triplet exciton, this annihilation effect provides a strong efficiency-limiting decay channel in these OLEDs.

Naturally, in contrast to the RISC process, TTA is a population density dependent process and thus scales with the injected current density. In order to assess this dependency, we carried out the same measurements and evaluation procedure for two further current densities: $5\frac{mA}{cm^2}$ and $20\frac{mA}{cm^2}$. The results reveal that the influence of TTA at room temperature is, as expected, increasing with current density. While at the lower current density of $5\frac{mA}{cm^2}$, 47% of the triplet excitons are depopulated by TTA at room temperature, this value increases to 55% for a current density of $20\frac{mA}{cm^2}$.

A quarter of the discussed TTA events produces singlet excitons (equation 1), which is represented as red, dotted area in Figure 5a. Reviewing equation 5, the singlet states are populated by this quarter of TTA events ($\frac{1}{4}k_{TT}T_1^2$) as well as triplet excitons upconverted by RISC ($k_{RISC}T_1$) and polarons that directly recombine to singlet excitons ($\frac{1}{4}\gamma n^2$). These processes are summarized in **Figure 5b**, illustrating the mentioned population channels (TTA, RISC, polarons) together with depopulation channels (singlet decay, ISC) of the singlet states. To investigate the influence of the different population processes of the singlet state by polarons, RISC and TTA on the EL, we analyze the amount of excitons present in the singlet state (gray box) initially populated by the mentioned processes (depopulation processes from the singlet states are included iteratively). In order to quantify the influence of these population processes on the EL, the diagram in Figure 5b illustrates the contributions of singlet excitons generated via polarons, RISC and TTA to the actual light emission (added up to 100%). Since more triplet excitons are lost with increasing temperature, caused by TTA (Figure 5a), less triplet excitons can be upconverted to the singlet state. Therefore, the percentage in the EL of singlet excitons formed directly by polarons increases with temperature. Analogously, as already shown in Figure 5a, the proportion of triplet excitons that are depopulated by RISC decreases with temperature. Accordingly, the number of singlet excitons created by RISC follows the same trend (the green part in Figure 5b also counts singlet excitons, which are formed from polarons, subsequently undergo ISC and are again upconverted by RISC). The interesting fact is that TTA events, whereof only 25% generate singlet excitons (equation 1), also contribute to the EL (red, dotted part). Since this path of producing singlet excitons requires twice as many triplet excitons as RISC and only takes place in a quarter of the occurring TTA events, it is by far a less efficient way of triplet harvesting.

CONCLUSION

We established a kinetic model for electrically driven TADF OLEDs including radiative and non-radiative first- and second-order effects that reproduces transient EL measurements accurately. We developed a suitable fit procedure to analyze the kinetic processes and to investigate the influence of efficiency-limiting effects in operational OLEDs. This method is especially advantageous for exciplex-based systems, since rate extraction with transient PL is considerably complicated by non-absorbing exciplex states and the associated requirement for initial optical excitation of the donor/acceptor molecules. In the established exciplex-based model system m-MTDATA:3TPYMB, we thereby revealed triplet-triplet annihilation (TTA) as the dominant second-order effect. Remarkably, TTA accounts for a significant part to triplet depopulation as a result of a moderate RISC rate and a high triplet density, especially in electrical excitation. Since one TTA event results in the loss of at least one triplet exciton, this is an important quantum efficiency-limiting loss channel for this type of OLEDs. We determined the influence of TTA increasing with temperature, which leads to TTA limiting the overall performance of OLEDs especially at room temperature. Consequently, the EL does not only consist of prompt and delayed constituents due to RISC, it also contains a contribution of emissive excited states formed via TTA. Since TTA is a less efficient triplet harvesting mechanism in contrast to RISC, these results represent an important insight for research on exciplex-based TADF OLEDs. We propose to use transient EL analysis with this adapted kinetic model as a standard tool to study the properties, possible efficiency-limiting processes and contributions to EL of electrically driven TADF OLEDs.

## ASSOCIATED CONTENT

**Supporting Information**

Supporting Information includes more details about the fit procedure and its python source code, all results including fitting errors, temperature-dependent EL spectra, voltage-dependent transient EL measurements, excitation power-dependent PL measurements, photoexcitation/-emission spectra, temperature dependence of TTA rate $k_{TT}$, temperature-dependent trEL measurements including fits, transformation of TTA rate in equation 1.

## AUTHOR INFORMATION

**Corresponding Author**

*Andreas Sperlich: sperlich@physik.uni-wuerzburg.de

**ORCID**

Jeannine Grüne: 0000-0003-3579-0455

Vladimir Dyakonov: 0000-0001-8725-9573

Andreas Sperlich: 0000-0002-0850-6757

**Author Contributions**

The manuscript was written through contributions of all authors. All authors have given approval to the final version of the manuscript.

**Notes**

The authors declare no competing financial interest.

ACKNOWLEDGMENT

J.G., V.D. and A.S. acknowledge the German Research Foundation, DFG, within GRK 2112. N.B. acknowledges the German Research Foundation, DFG, within the SPP 1601 (SP1563/1-1). A.S. acknowledges EU H2020-MSCA-ITN-2016 "SEPOMO" 722651. We thank Dr. Kristofer Tvingstedt and Mathias Fischer from University of Würzburg for useful discussions. Furthermore, we thank Philipp Rieder from University of Würzburg for measurements of temperature-dependent EL spectra.

**TOC Graphic**

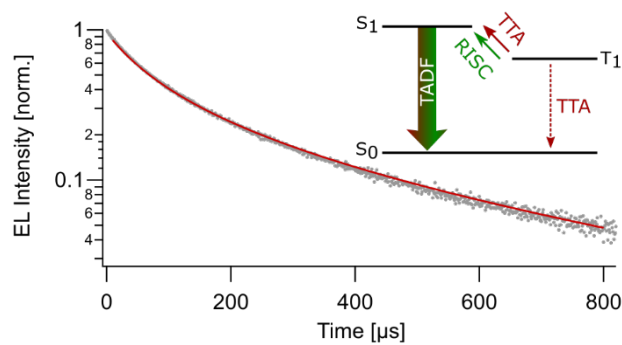

# Kinetic Modeling of Transient Electroluminescence reveals TTA as Efficiency-Limiting Process in Exciplex-Based TADF OLEDs

Jeannine Grüne[1], Nikolai Bunzmann[1], Moritz Meinecke[1],
Vladimir Dyakonov[1], Andreas Sperlich[1*]

[1]Experimental Physics 6, Julius Maximilian University of Würzburg, Am Hubland, 97074 Würzburg, Germany

**Fit Results and Fit Procedure**

**Table 1.** Temperature-dependent fit results (± fit error) for m-MTDATA:3TPYMB OLEDs. The measurements are performed at a current density of $10\,\frac{mA}{cm^2}$ and the singlet decay rate $k_S$ was taken from literature (see below).

| $T$ [K] | $k_{TT}$ [$10^{-14}$ cm$^3$s$^{-1}$] | $k_T$ [$10^3$ s$^{-1}$] | $k_{RISC}$ [$10^4$ s$^{-1}$] | $k_{ISC}$ [$10^6$ s$^{-1}$] | $k_S$ [$10^5$ s$^{-1}$] | $j$ [mA cm$^{-2}$] |
|---|---|---|---|---|---|---|
| 200 | 1.432 ± 0.015 | 0.1 [1] | 4.7 | 3.4 [2] | 1.0 | 10 |
| 220 | 3.706 ± 0.011 | 2.3 | 5.3 [3] | 3.4 | 1.0 | 10 |
| 240 | 9.995 ± 0.027 | 4.846 ± 0.023 | 5.9 | 3.4 | 1.0 | 10 |
| 260 | 19.766 ± 0.097 | 8.368 ± 0.061 | 6.4 | 3.4 | 1.0 | 10 |
| 280 | 40.07 ± 0.24 | 9.299 ± 0.10 | 6.9 | 3.4 | 1.0 | 10 |
| 300 | 67.29 ± 0.50 | 11.910 ± 0.16 | 7.4 | 3.4 | 1.0 | 10 |

[1] Fit with boundary conditions for 200 K and 220 K ($k \geq 0$) → no realistic error computable for $k_T$, $k_{RISC}$ and $k_{ISC}$

[2] $k_{ISC}$ temperature-independent[1]

[3] $\Delta E_{ST}$ = 22.7 meV follows from fitting $k_{RISC}$ at 220 K (see below)

The RISC rate $k_{RISC}$ was determined for 200 K because this temperature provides the longest transient and therefore the smallest fit inaccuracy. The singlet-triplet gap $\Delta E_{ST}$ was determined by fitting the transient EL measurement at 220 K and assuming a Boltzmann-activated process from The RISC rate from 200 K. Since the rates to be fitted are not independent from each other, the knowledge of an exponentially increasing RISC rate with the Boltzmann term was used. For the transients from 240 K – 300 K, the RISC rates were calculated with the RISC rate for 200 K considering the fitted energy gap $\Delta E_{ST}$ of 22.7 meV:

$$k_{RISC,2x0K} = k_{RISC,200K} \cdot \exp\left(\left(\frac{1}{T_{200K}} - \frac{1}{T_{2x0K}}\right) \cdot \frac{\Delta E_{ST}}{k_B}\right)$$

---

* sperlich@physik.uni-wuerzburg.de

This implies that for the transients for 240 K - 300 K, there were only two free fitting parameters $k_T$ and $k_{TT}$. The errors indicated in Table 1 correspond to the fit errors. The decay rate $k_S$ was taken from literature.[2] The Langevin recombination rate was determined with $\varepsilon_r = 2.9$[3,4], $\mu_e = 10^{-5} \frac{cm^2}{Vs}$ and $\mu_h = 10^{-4} \frac{cm^2}{Vs}$ [5,6]. The fits for all temperatures are displayed in Figure S6.

**Fitting transient EL data with Python:**

For fitting the transient EL measurements, we used an iterative procedure (forward Euler method), in which the rate equations are solved by stepwise integration with a timestep Δt, small enough to properly model the data. The fitting was realized by a standard Levenberg-Marquardt-Algorithm, implemented in the *Scipy* library (v. 1.1.0) in Python (v. 3.7.6), using the simulated rate equation model as fitting function. The script used for the iterative solution of the rate equations including calculation of loss processes and contribution to EL looks as follows:

```python
def trEL(x,kr,knr,kT,kisc,krisc,kTP,kTT,kST,kSP,j,T,E):

    y = 6.86e-11 # epsilon_r=2.9 (in cm)
    e = 1.6e-19  # elementary charge
    d= 70e-7     # OLED emission layer thickness (in cm)
    dt = 0.0000001 # time step

    n = int(0.01/dt)    # steps in simulated time interval
    n1 = int(n/2)       # turn-off point

    # State vectors
    S1 = np.zeros(n)
    T1 = np.zeros(n)
    m = np.zeros(n)
    tim = np.zeros(n)
    tim[0]=-n1*dt
    EL = np.zeros(n)
    dEL = np.zeros(n)
    IQE = np.zeros(n)
    # Loss processes
    TTA_loss = np.zeros(n)
    TTA_sing = np.zeros(n)
    RISC_loss = np.zeros(n)
    kT_loss = np.zeros(n)
    # Normalized Loss processes
    n_TTA = np.zeros(n)
    n_RISC = np.zeros(n)
    n_pol = np.zeros(n)
    n_TTA_norm = np.zeros(n)
    n_RISC_norm = np.zeros(n)
    n_pol_norm = np.zeros(n)

    # krisc is increasing with Boltzmann factor:
    z = ((1/200)-(1/T))*(1/8.61733e-5)
    krisc = krisc*math.exp(z*E)

    # Simulation
    for i in range(1,n):
        if i<n1:
            tim[i]=tim[i-1]+dt
            m[i] = m[i-1] + (j/(e*d) - y*m[i-1]*m[i-1])*dt
            S1[i]=S1[i-1]+(-(kr+knr)*S1[i-1]+krisc*T1[i-1]-kisc*S1[i-1]
                    +(1/4)*kTT*T1[i-1]*T1[i-1]-kST*T1[i-1]*S1[i-1]
                    -2*kSP*m[i-1]*S1[i-1]+0.25*y*m[i-1]*m[i-1])*dt
            T1[i]=T1[i-1]+(-kT*T1[i-1]-krisc*T1[i-1]+kisc*S1[i-1]-2*kTP*m[i-1]*T1[i-1]
                    -(5/4)*kTT*T1[i-1]*T1[i-1]+0.75*y*m[i-1]*m[i-1])*dt
            EL[i]=kr*S1[i]
```

```
            #Loss processes of Triplet States
            TTA_loss[i]=(5/4)*kTT*T1[i]*T1[i]
            TTA_sing[i]=(1/4)*kTT*T1[i]*T1[i]
            RISC_loss[i]=krisc*T1[i]
            kT_loss[i]=kT*T1[i]
            #Contribution to EL
            n_TTA[i]=n_TTA[i-1]+((1/4)*kTT*T1[i-1]*T1[i-1]-(kr+knr+kisc)*n_TTA[i-1])*dt
            n_RISC[i]=n_RISC[i-1]+(krisc*T1[i-1]-(kr+knr+kisc)*n_RISC[i-1])*dt
            n_pol[i]=n_pol[i-1]+(0.25*y*m[i-1]*m[i-1]-(kr+knr+kisc)*n_pol[i-1])*dt
        if n1<=i and i<n:    #j=0 to receive transient EL decay
            tim[i]=tim[i-1]+dt
            m[i]= m[i-1] +(- y*m[i-1]*m[i-1])*dt
            S1[i]=S1[i-1]+(-(kr+knr)*S1[i-1]+krisc*T1[i-1]-kisc*S1[i-1]
                    +(1/4)*kTT*T1[i-1]*T1[i-1]-kST*T1[i-1]*S1[i-1]
                    -2*kSP*m[i-1]*S1[i-1]+0.25*y*m[i-1]*m[i-1])*dt
            T1[i]=T1[i-1]+(-kT*T1[i-1]-krisc*T1[i-1]+kisc*S1[i-1]-2*kTP*m[i-1]*T1[i-1]
                    -(5/4)*kTT*T1[i-1]*T1[i-1]+0.75*y*m[i-1]*m[i-1])*dt
            EL[i]=kr*S1[i]
            #Loss processes of Triplet States
            TTA_loss[i]=(5/4)*kTT*T1[i]*T1[i]
            TTA_sing[i]=(1/4)*kTT*T1[i]*T1[i]
            RISC_loss[i]=krisc*T1[i]
            kT_loss[i]=kT*T1[i]
            #Contribution to EL
            n_TTA[i]=n_TTA[i-1]+((1/4)*kTT*T1[i-1]*T1[i-1]-(kr+knr+kisc)*n_TTA[i-1])*dt
            n_RISC[i]=n_RISC[i-1]+(krisc*T1[i-1]-(kr+knr+kisc)*n_RISC[i-1])*dt
            n_pol[i]=n_pol[i-1]+(0.25*y*m[i-1]*m[i-1]-(kr+knr+kisc)*n_pol[i-1])*dt

    # normalize contribution to EL on steady-state singlet population
    n_TTA_norm = n_TTA/(S1[n1-100])*100
    n_RISC_norm = n_RISC/(S1[n1-100])*100
    n_pol_norm = n_pol/(S1[n1-100])*100

    # normalize simulation
    for i in range(1,n):
        dEL[i]=EL[i]/EL[n1-100]

    # interpolate dEL output on x-vector (to allow fitting)
    dEL_out = np.interp(x,tim,dEL)

    # Outputs
    return dEL_out, tim, TTA_loss, TTA_sing, RISC_loss, kT_loss, n_TTA, n_RISC, n_pol,
            n_TTA_norm, n_RISC_norm, n_pol_norm
```

The defined function `trEL` is fitted using `optimize.curve_fit()` from the *Scipy* library to the imported transient EL data. As described above, output parameters from the fit are for 200 K: $k_{\text{RISC}}$, $k_{\text{ISC}}$, $k_{\text{T}}$ and $k_{\text{TT}}$, for 220 K: $E$, $k_{\text{T}}$ and $k_{\text{TT}}$, and for 240 – 300 K: $k_{\text{T}}$ and $k_{\text{TT}}$. With the given fit results, the loss processes are calculated as depopulated triplet density per second and the contribution to EL in percentage. Subsequently, the loss processes are normalized to the sum of `TTA_loss`, `RISC_loss` and `kT_loss`.

# Figures

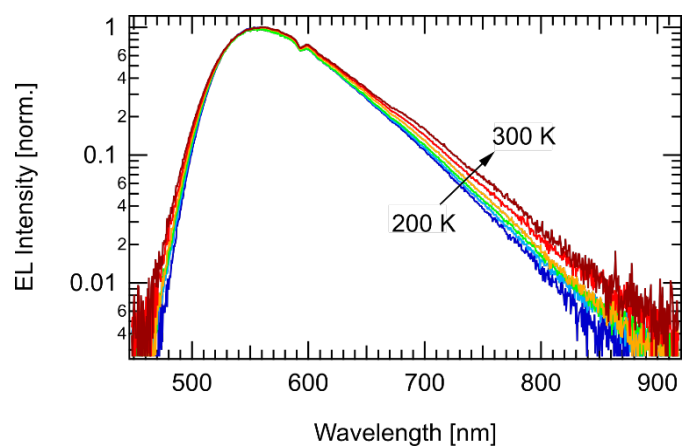

**Figure S1.** Temperature-dependent EL spectra in a range from 200 K to 300 K (20 K steps). The EL peak does not shift with temperature, but a slight broadening can be observed. All spectra show a sinusoidal feature at 590 nm to 600 nm, which is an artefact related to attenuation of the spectrometer's multimodal glass fiber.

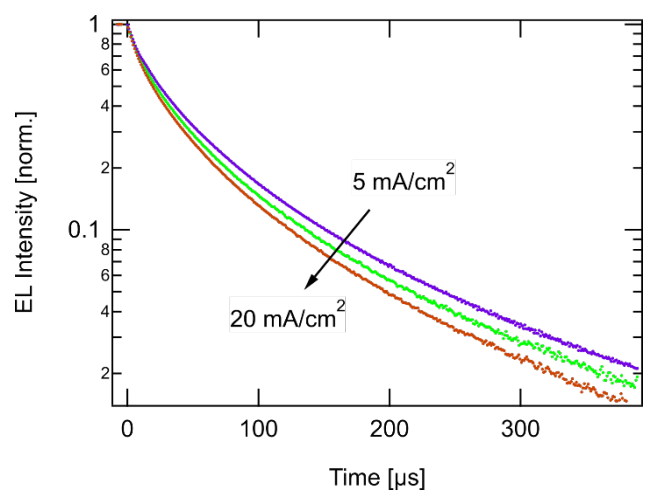

**Figure S2.** Current-dependent transient EL measurements at 260 K. The transient EL decay becomes faster with increasing applied current density.

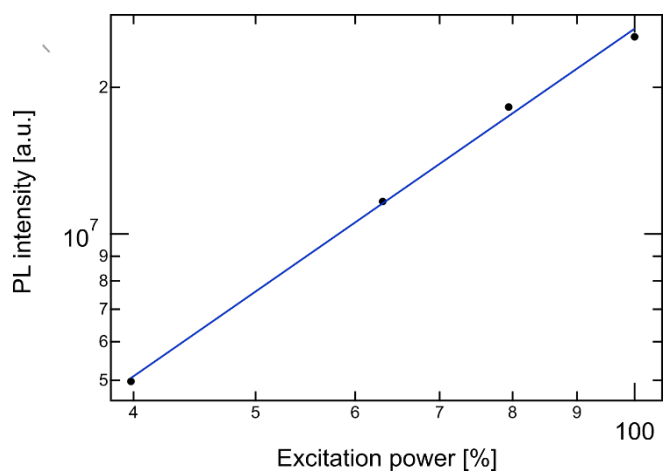

**Figure S3.** Excitation power-dependent PL intensity of a solid film of m-MTDATA:3TPYMB blend in log-log representation. The PL intensity increases with a slope of 1.8 ± 0.1 (fit in blue), indicating a second-order process.

Experimental Details of this experiment: PL spectra of a m-MTDATA:3TPYMB blend (20 mg/ml in chlorobenzene spin-coated on a glass substrate) were recorded with a FLS980 spectrometer from Edinburgh Instruments. The excitation source was a Xenon Lamp with different transmission filters (OD 0.1, OD 0.2 and OD 0.4) and a selected emission wavelength of 365 nm. Possible transmissions of the xenon lamp or fluorescence of the glass substrate were subtracted by separately measured spectra of the pure glass substrate.

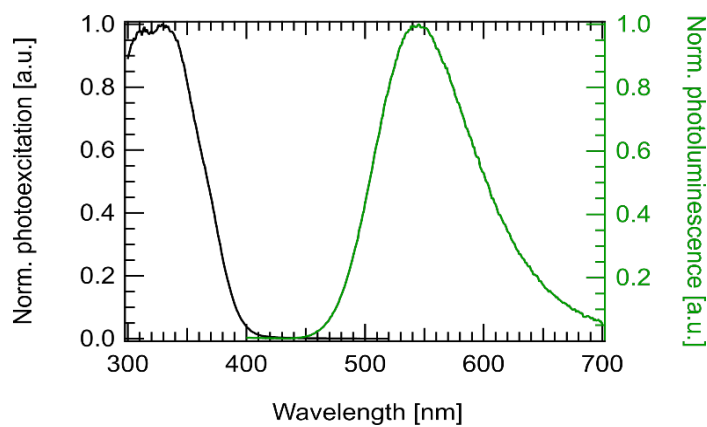

**Figure S4:** Photoexcitation (black) and photoluminescence (green) spectra of solid films of m-MTDATA:3TPYMB blend. The CT/exciplex state shows no absorption.

Experimental details of this experiment: Photoexcitation (PE) and PL spectra of a solid film of m-MTDATA:3TPYMB blend (spin-coated from 20 mg/ml chlorobenzene solution) were recorded with a FLS980 spectrometer from Edinburgh Instruments. Photoexcitation was measured at an emission wavelength of 540 nm. Photoluminescence was measured with an excitation wavelength of 365 nm.

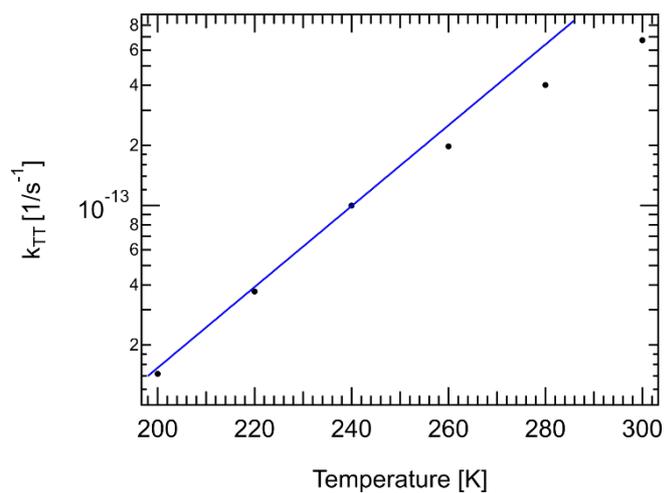

**Figure S5.** Temperature dependence of the determined TTA rate $k_{TT}$. For lower Temperatures, the rate is exponentially temperature-activated following $k_{TT} \sim exp\,(const \cdot T)$ with $const = (46.6 \pm 2.6)10^{-3}K^{-1}$ (fit in blue). For higher temperatures, the slope is flattening. Nevertheless, in contrast to the RISC rate, the temperature-activation is significantly more pronounced for the TTA rate.

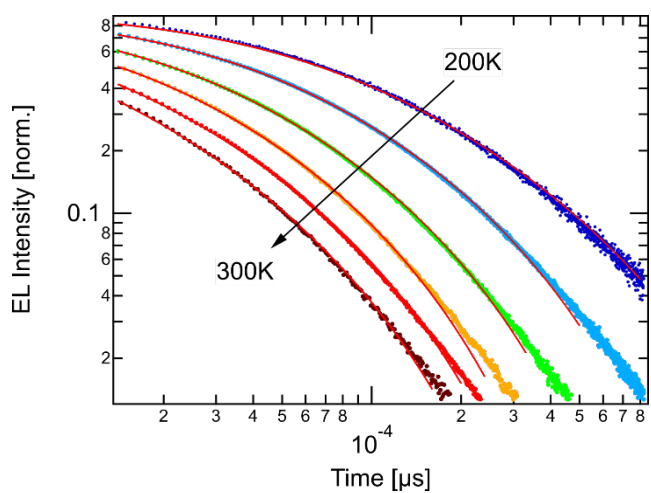

**Figure S6.** Fit results for transient EL measurements from 200 K to 300 K. The fit is always starting at 10 μs to avoid the influence of electrical artifacts. The fit can deviate at the end of the curve, since the EL intensity is not as strong there and is not weighted as much in the fit.

**Transformation of TTA rate (equation 1):**

The TTA process is split up into three parts, either forming a quintet, triplet or the singlet state (equation 1). Hence, the individual loss-rates with respect to the original triplet population are given by the probability of the transition multiplied with the number of lost triplets in the process. This results in rates of $\frac{5}{9} k'_{\mathrm{TT}}$ for the depopulation of the triplet state and $\frac{1}{9} k'_{\mathrm{TT}}$ for the population of the singlet state through TTA. Defining $k_{TT} = \frac{9}{4} k'_{\mathrm{TT}}$, these rates can be transformed to the following:

$$\frac{dT_1}{dt} = \ldots - \left(0 \cdot \frac{5}{9} + 1 \cdot \frac{3}{9} + 2 \cdot \frac{1}{9}\right) k'_{\mathrm{TT}} = \ldots - \frac{5}{9} k'_{\mathrm{TT}} =^* \ldots - \frac{5}{4} k_{\mathrm{TT}}$$

$$\frac{dS_1}{dt} = \ldots + \left(1 \cdot \frac{1}{9}\right) k'_{\mathrm{TT}} = \ldots + \frac{1}{9} k'_{\mathrm{TT}} =^* \ldots + \frac{1}{4} k_{\mathrm{TT}}$$

*) $k_{\mathrm{TT}} = \frac{9}{4} k'_{\mathrm{TT}}$

In context of TTA, the formation of a quintet state is sometimes not considered at all. In this way the given rates are formed directly:

$$\frac{dT_1}{dt} = \ldots - \left(1 \cdot \frac{3}{4} + 2 \cdot \frac{1}{4}\right) k_{\mathrm{TT}} = \ldots - \frac{5}{4} k_{\mathrm{TT}}$$

$$\frac{dS_1}{dt} = \ldots + \left(1 \cdot \frac{1}{4}\right) k_{\mathrm{TT}} = \ldots + \frac{1}{4} k_{\mathrm{TT}}$$